\documentclass[fleqn,usenatbib]{mnras}

\usepackage{newtxtext,newtxmath}

\usepackage[T1]{fontenc}

\DeclareRobustCommand{\VAN}[3]{#2}
\let\VANthebibliography\thebibliography
\def\thebibliography{\DeclareRobustCommand{\VAN}[3]{##3}\VANthebibliography}

\usepackage{graphicx}	
\usepackage{amsmath}	
\usepackage[normalem]{ulem}
\usepackage[dvipsnames,table]{xcolor}

\newcommand{\msun}{{\,\rm M_\odot}}

\newcommand{\kms}{\,{\rm km}\,{\rm s}^{-1}}

\def\gsim{ \lower .75ex \hbox{$\sim$} \llap{\raise .27ex \hbox{$>$}} }
\def\lsim{ \lower .75ex \hbox{$\sim$} \llap{\raise .27ex \hbox{$<$}} }

\title[CDM and WDM at JWST frontier]{Halo assembly in cold and warm dark matter during the JWST frontier epoch}

\author[M.~R.~Lovell]{
Mark R. Lovell$^{1}$\thanks{E-mail: lovell@hi.is}
\\
$^{1}$Centre for Astrophysics and Cosmology, Science Institute, University of Iceland, Dunhaga 5, 107 Reykjav\'ik, Iceland}

\date{Accepted XXX. Received YYY; in original form ZZZ}

\pubyear{2023}

\begin{document}
\label{firstpage}
\pagerange{\pageref{firstpage}--\pageref{lastpage}}
\maketitle

\begin{abstract}
The {\it JWST} mission is in the process of probing the galaxy mass function at $z>10$, when conceivably any delay in halo assembly due to the presence of a dwarf galaxy-scale power spectrum cutoff may drastically suppress the number of galaxies relative to the cold dark matter (CDM) expectation. We employ $N$-body simulations of CDM and warm dark matter (WDM) to explore how the difference in halo collapse time between these models scales with $z=0$ descendant halo mass. We demonstrate that collapse begins first for the most massive haloes, and the delay in collapse time between CDM and WDM haloes correlates inversely with descendant mass. We thus infer that only present-day dwarf galaxies exhibit any difference in their assembly history between CDM and WDM at $z=10$, and therefore support previous studies that have found {\it JWST} is unlikely to determine whether our Universe is better described by the CDM cosmology or the WDM cosmology without favourable lensing studies.
\end{abstract}

\begin{keywords}
dark matter -- galaxies: high redshift
\end{keywords}



\section{Introduction}

The clustering properties of dark matter on scales $\gsim\rmn{Mpc}$ are constrained strongly in the present day Universe through observations such as baryon acoustic oscillations \citep{Eisenstein05}. These large scale constraints point to dark matter that has negligible self-interactions \citep{Robertson19} and negligible velocity dispersion in the early Universe \citep{White83}. A dark matter model with this combination of properties is known as cold dark matter (CDM). Attempts to better distinguish between CDM and the warm dark matter (WDM) model, the latter of which relaxes the assumption of negligible early-Universe velocity dispersion, typically focus on dwarf galaxies in the Local Group \citep[LG,][]{Lovell17a,Lovell17b,Kim17,Newton21,Nadler21}. The primary advantage of the LG as a dark matter physics laboratory is that it is the one regime in which the lowest luminosity galaxies can be detected and analysed, which is the class of galaxies where the degeneracy between baryon physics and novel dark matter physics is weakest \citep{di-cintio2014}.

There are two important disadvantages to the LG option. First, LG galaxies constitute a very small sample of the whole Universe, so the impact of the cosmic variance on the statistics can be substantial. Secondly, the low-redshift dwarf galaxy population is a product of a very non-linear and partially stochastic in nature galaxy formation mechanisms \citep[e.g.][]{Cole00,Vogelsberger14,Schaye15}.
An alternative set of opportunities presents itself at high redshift, where the absence of low mass galaxies in models with a matter power spectrum cutoff leads to a reduction in ionizing photons, and thus delays the completion of reionization compared to CDM \citep{Bose16c,Menci16,Lovell18a,VillanuevaDomingo18,Dayal23}. 

One further consequence of alternative dark matter models at high redshift is the delay in assembly time of low mass haloes that form around the cutoff \citep{Lovell12,Maio15,Maccio19,Esmerian21}, which in principle exacerbates the difference between CDM and WDM at very high redshifts. Such a difference could be detected with a survey of sufficient depth and efficiency. It is also the case that the temperature at which gas can cool in haloes is lowest at early times \citep{Benson_02,Bullock_00,BenitezLlambay20}; therefore high-redshift low mass haloes are more likely to be luminous than their low redshift equivalents and thus may in some ways be easier to detect. Several studies have used hydrodynamical simulations to discern whether the {\it JWST} mission is capable of discriminating between dark matter models that do and do not have a cutoff at the mass scale of $\sim10^{8}$~$\msun$ \citep{Maio15,Bose16c,Rudakovskyi16,Lovell18a,Kurmus22,Maio23,Shen23}. All of these studies predict that the number of detectable galaxies at $z=10-12$ with {\it JWST} is the same in models with and without a cutoff, with strong magnification by gravitational lensing to have any chance of discerning between the models. 

We can obtain a first order understanding of this result through consideration of the spherical matter perturbation turnaround time, $t_\rmn{t}$, which is approximately the time at which the perturbation decouples from the Hubble flow and starts the process of forming a stable, virialised halo. It can be inferred from the parametric solution to the gravitational orbit of a test particle in such a perturbation that:

\begin{equation}
    t_\rmn{t}\propto \delta_{0}^{-3/2},
    \label{eqn:td0}
\end{equation}

\noindent where $\delta_{0}$ is the overdensity the spherical perturbation would have at the present day if it only ever underwent linear growth; see \citet{Eke96} for a more careful treatment. The highest peaks in $\delta_{0}$ will occur where a series of perturbations are stacked on top of one another: for example, a dwarf-galaxy scale perturbation within a massive galaxy perturbation that itself is within a cluster perturbation. The removal of any of the perturbations in the stack will lower $\delta_{0}$ and thus delay the turnaround time. If we compare the evolution of two $z=0$ haloes with identical masses, one of which is located in an overdense region and the second in an underdense environment, the absence of a large scale perturbation in the underdense region will therefore lead to a later collapse time for the second halo. Accordingly, galaxy-scale perturbations embedded in large-scale underdense regions, like cosmic voids, will likewise exhibit retarded collapse times. On the other hand, WDM suppresses small scale perturbations, especially at mass scales around and below the WDM half-mode mass scale, $M_\rmn{hm}$. Therefore, as with the absence of large scale perturbations in underdense regions, the highest density peak value of $\delta_{0}$ within the perturbation is lower than in CDM and thus the collapse occurs later than it would if there were perturbations present of mass $<M_\rmn{hm}$.

The delay with respect to CDM will be most dramatic for haloes where the suppression of $\delta_{0}$ by free-streaming is largest, which will be haloes that satisfy the following two conditions: (i) their masses are only marginally larger than $M_\rmn{hm}$, and (ii) they are not superimposed on any part of a more massive (i.e. larger) density fluctuation. Therefore, haloes much more massive than $M_\rmn{hm}$ will receive little to no delay in WDM relative to CDM. The first progenitors of these haloes in WDM will have masses of order $\sim M_\rmn{hm}$, yet the large local overdensity will ensure the collapse time is similar to that of CDM. Some of these progenitors may survive as subhaloes to the present day, and therefore we would expect that their collapse delay relative to CDM will be shorter than for isolated haloes in the field. 

A more complete treatment of this topic will involve details of extended-Press-Schechter (EPS) theory \citep{Press74,Bond91}.
The distribution of small perturbations within larger perturbations prior to non-linear collapse, and the evolution of both the most massive progenitor and the collapsed fraction across cosmic time, was previously analysed in the COCO simulations by \citet{Ludlow16}. In our study we instead perform an abbreviated analysis with a specific focus on $z\ge10$, which we term the {\it JWST frontier epoch} in recognition of how this facility opens up a crucial new window above and beyond previous observatories. In particular, we identify $z>10$ progenitors of present day haloes, measure the mass present in progenitors and collapse delay as a function of present day mass, and observe how collapse time changes for $z=0$ dwarf haloes as a function of their host mass.

In this paper we examine the dark matter halo assembly in the {\it JWST frontier epoch}. We use the COCO $N$-body simulations of CDM and WDM (3.3~keV thermal relic) to demonstrate how the collapse time--overdensity relation discussed above affects the mass assembly of present day objects. This paper is organised as follows: in Section~\ref{sec:sims} we summarise the simulations used in this paper, and in Section~\ref{sec:res} we present our results. We draw conclusions in Section~\ref{sec:conc}.   

\section{Simulations}
\label{sec:sims}

The two simulations used in this project are the $N$-body {\it Copernicus Complexio} (COCO) simulations. The COCO volume is a zoomed $\sim$28~Mpc radius sphere set within a parent box of side-length 100~Mpc. The simulation particle mass is $1.93\times10^{5}$~$\msun$ in the high resolution region, and the gravitational softening length $\epsilon=0.33$~kpc. The cosmological parameters are selected to be consistent with the {\it WMAP-7} results \citep{wmap11} and their values are:  Hubble parameter $h=0.704$, matter density $\Omega_{0}=0.272$, dark energy density, $\Omega_{\Lambda}=0.728$; power spectrum normalization $\sigma_{8} = 0.81$ and spectral index, $n_\rmn{s}=0.967$. One of the simulations adopts the CDM linear matter power spectrum \citep{Hellwing16}, and the other uses a WDM linear matter power spectrum \citep{Bose16a}. The WDM particle is a 3.3~keV thermal relic, which corresponds to a half-mode mass $M_\rmn{hm}=3.5\times10^{8}$~$\msun$. This parameter was chosen to be in mild tension with the Lyman-$\alpha$ forest constraints derived by \citet{Viel13}. Finally, haloes are identified using the geometric friends-of-friends (FoF) algorithm, and are decomposed into gravitationally bound smooth haloes and subhaloes using the {\sc subfind} algorithm \citep{Springel01}. A minimum of 20 particles is required to identify a halo or subhalo.

\begin{figure*}
    \centering
      \setbox1=\hbox{\includegraphics[scale=0.23]{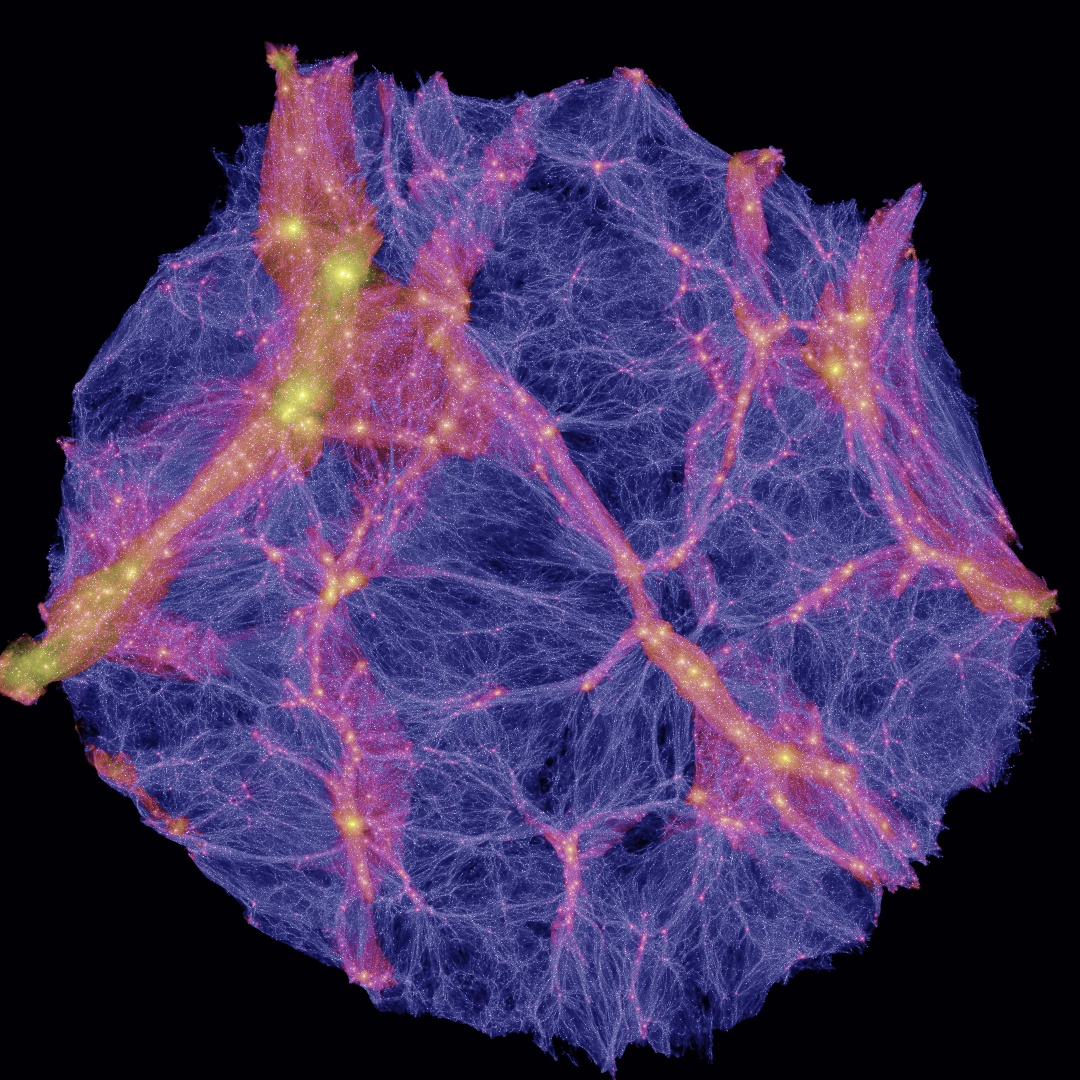}}
    \includegraphics[scale=0.23]{Figures/COCO_CDM_159_Bsize20000kpc_Sthick1000kpc_ImSetAll.jpg}\llap{\makebox[1.05\wd1][l]{\raisebox{-0.1\wd1}{\includegraphics[width=0.36\textwidth]{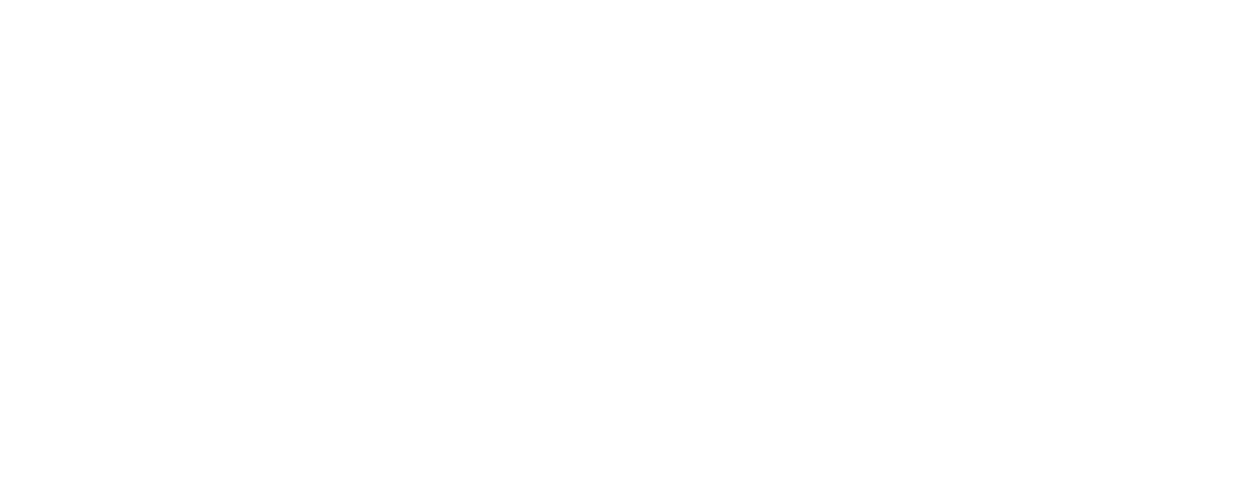}}}}\llap{\makebox[0.3\wd1][l]{\raisebox{-0.1\wd1}{\includegraphics[width=0.36\textwidth]{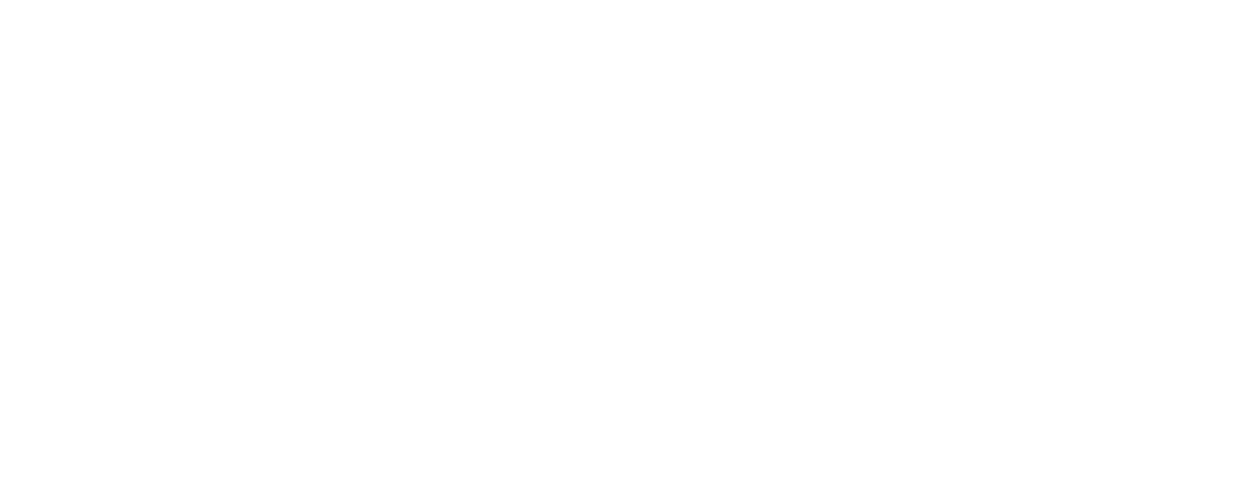}}}}
    \includegraphics[scale=0.23]{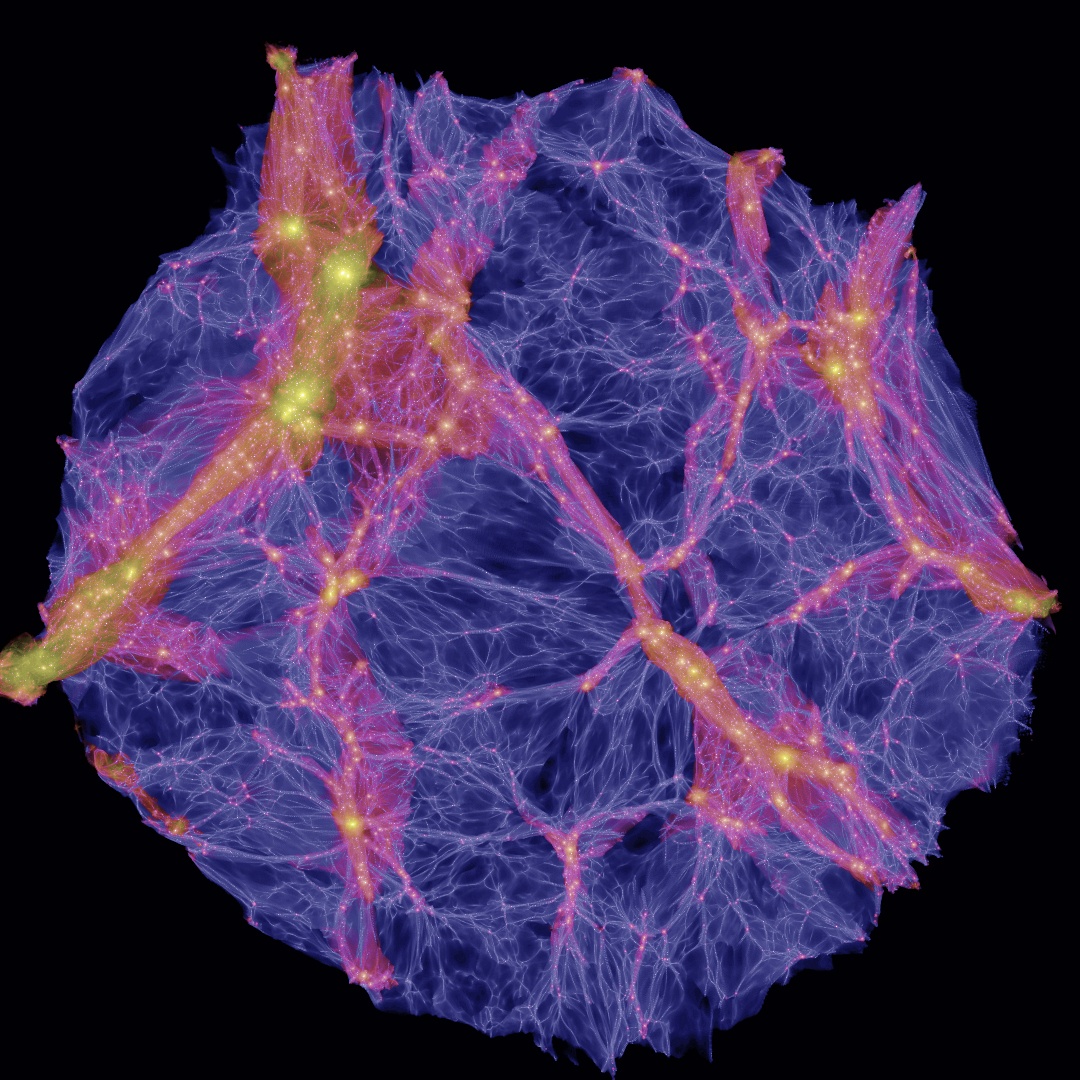}\llap{\makebox[1.05\wd1][l]{\raisebox{-0.1\wd1}{\includegraphics[width=0.36\textwidth]{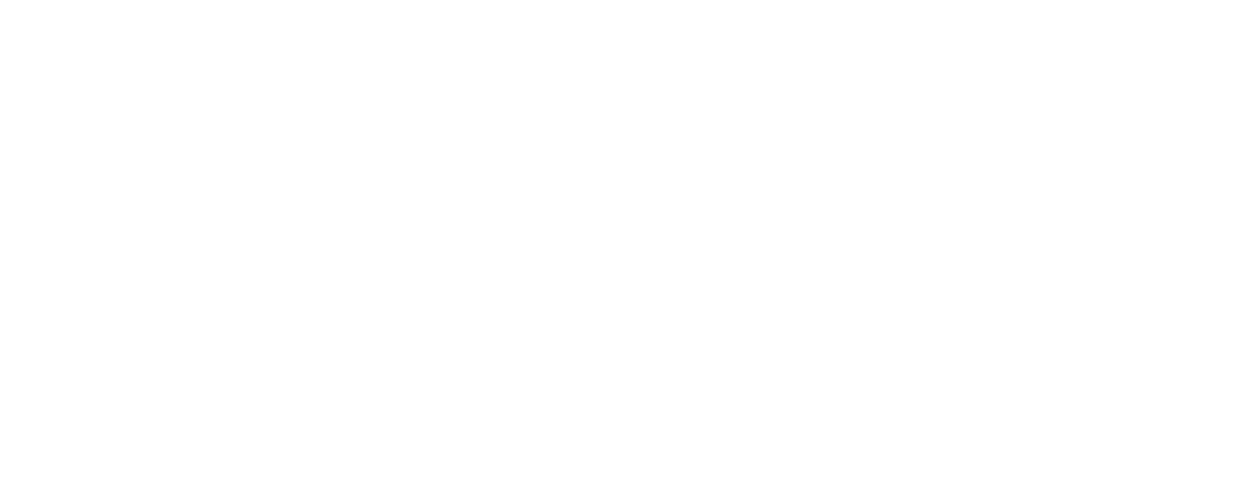}}}}\llap{\makebox[0.3\wd1][l]{\raisebox{-0.1\wd1}{\includegraphics[width=0.36\textwidth]{Figures/COCO_Redshift0p0_Label-eps-converted-to.pdf}}}}
    \includegraphics[scale=0.23]{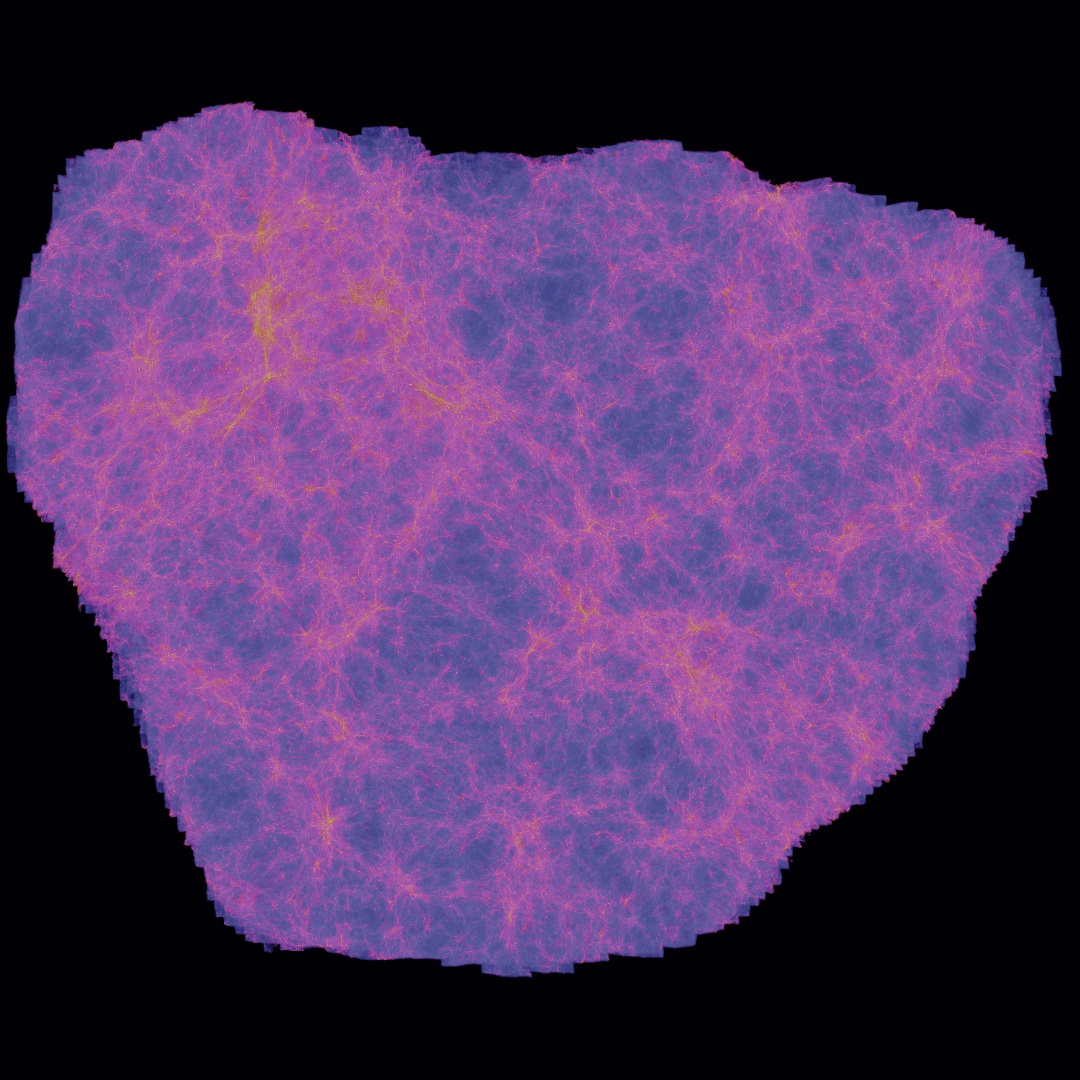}\llap{\makebox[1.05\wd1][l]{\raisebox{-0.1\wd1}{\includegraphics[width=0.36\textwidth]{Figures/COCO_CDM_Label-eps-converted-to.pdf}}}}\llap{\makebox[0.3\wd1][l]{\raisebox{-0.1\wd1}{\includegraphics[width=0.36\textwidth]{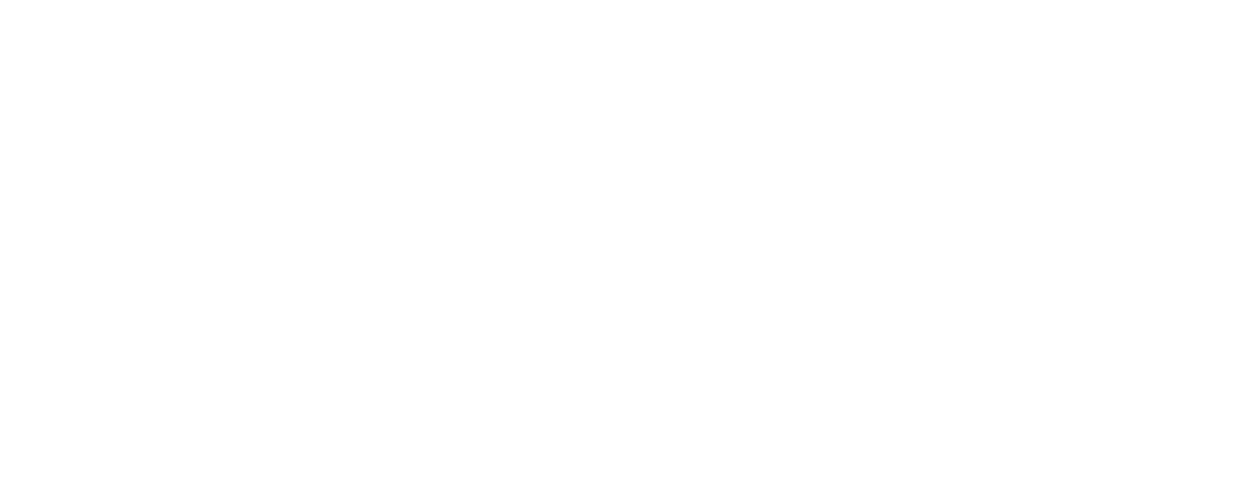}}}}
    \includegraphics[scale=0.23]{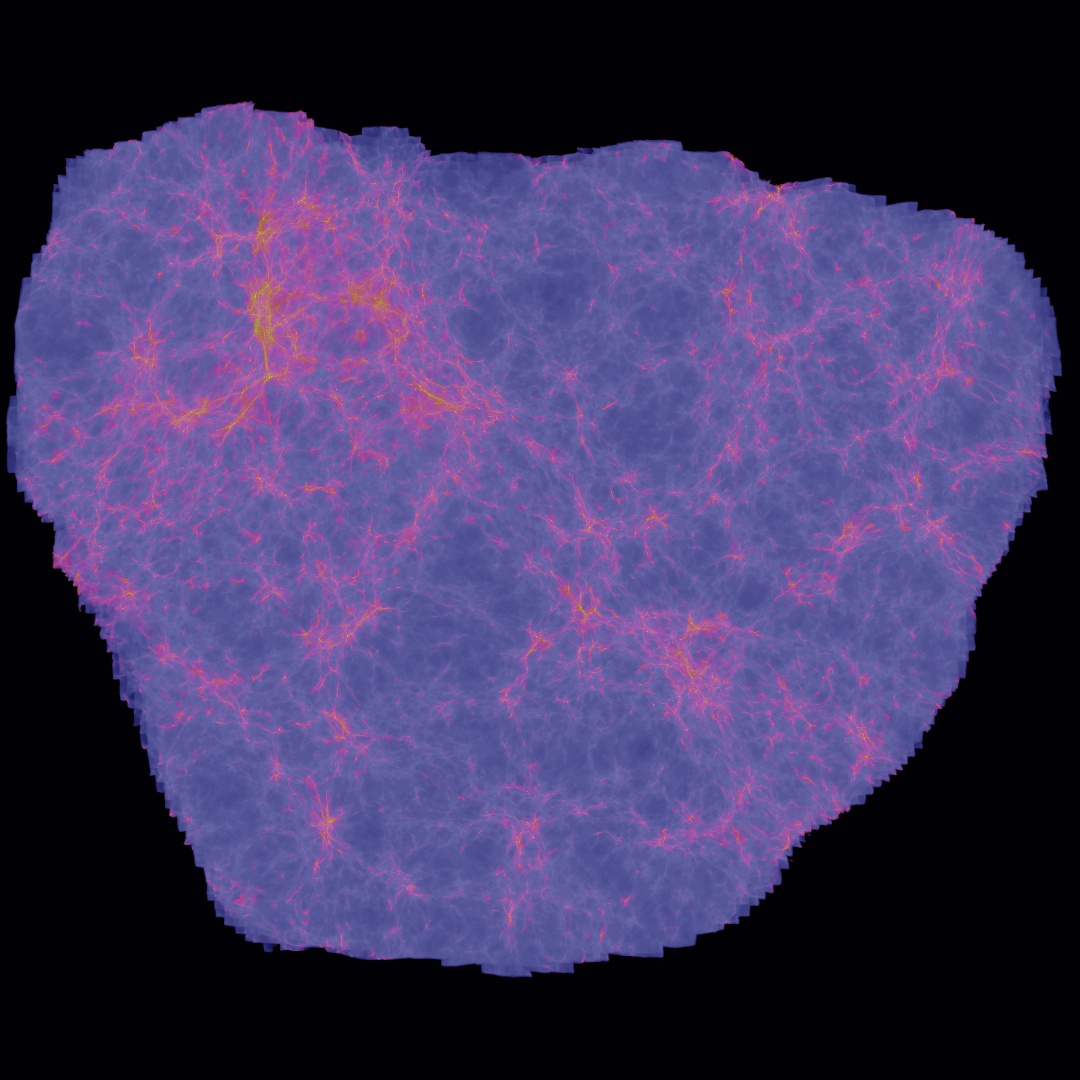}\llap{\makebox[1.05\wd1][l]{\raisebox{-0.1\wd1}{\includegraphics[width=0.36\textwidth]{Figures/COCO_WDM_Label-eps-converted-to.pdf}}}}\llap{\makebox[0.3\wd1][l]{\raisebox{-0.1\wd1}{\includegraphics[width=0.36\textwidth]{Figures/COCO_Redshift10p0_Label-eps-converted-to.pdf}}}}
    \caption{Images of the zoomed region in the CDM (left) and WDM (right) COCO simulations at $z=0$ (top panels) and $z=10$ (bottom panels). Each image consists of a slice 2.9~Mpc thick, or 1.45~Mpc either side of the high-resolution region centre, and is 57~Mpc on a side. Image intensity indicates dark matter comoving density and colour indicates velocity dispersion, with blue denoting low velocity dispersion ($<5$~$\kms$) and yellow high velocity dispersion ($>200$~$\kms$).}
    \label{fig:sims}
\end{figure*}

 In order to illustrate how WDM preserves the distribution of matter on large scales but radically smooths out perturbations on small scales, we generate high resolution CDM and WDM simulation images that cover the entire zoomed region. This region is 57~Mpc in diameter: we make images of a slice 2.9~Mpc thick that is centred in the middle of the sphere and present these images for both $z=0$ and $z=10$ in Fig.~\ref{fig:sims}. The image intensity indicates the projected density and the colour reflects the projected average three-dimensional velocity dispersion. 
 
 The locations of the massive haloes are identical at $z=0$ in the two models, as is the structure of the series of filaments that join them together. We present this result as the first indication that the formation time of these massive objects is practically identical in the two models. A dramatic delay would lead to collapse occurring at an epoch when the universe was much less dense and would conceivably lead to large differences in the appearance of structures. Such an effect is evident in WDM simulations where the particle mass is too low to produce even the detected number of Milky Way (MW) satellites \citep{Libeskind13}. Where the two do differ is in the intense speckling of low mass haloes in CDM: these haloes are entirely absent in WDM.  

We support this finding with the $z=10$ images shown in the lower panels. The spatial distribution of filaments and voids is again identical in the two models. The difference in low-mass halo abundance is much less apparent than at $z=0$; instead, the most striking difference between the two images is in the velocity dispersion. The large spatial distribution of medium-velocity dispersion in CDM, shown in pink, indicates where the material is collapsing into virialised objects: the relative predominance of blue colours for the  WDM structures reflects where this collapse has not taken place to the same degree in this model.

Our primary goal is to examine the properties of $z=0$ isolated haloes' and dwarf subhaloes' progenitors at redshifts $z\ge 10$, where we define isolated haloes as the most massive self-bound halo in each FoF group and all other haloes as subhaloes. Studies of halo evolution typically involve building merger trees between halo catalogues at subsequent snapshots and are interested in tracking halo assembly up to and around the peak of cosmic star formation, at around $z\sim2$ \citep[e.g.][]{Madau14}; the EAGLE model is only considered valid for $z<4$ \citep{Crain15}. However, the accuracy of the merger tree code and the subhalo finder are not sufficient to enable the tracking of haloes reliably from the present day all the way back to within 250~Myr after the Big Bang.  Therefore, we adopt a method based on particle IDs to identify links between $z\ge10$ progenitors and $z=0$ descendants. Given that isolated haloes gain mass in that time whereas subhaloes will lose some of their mass through stripping, we use slightly different algorithms for these two classes of object, and describe these algorithms below. For both isolated haloes and subhaloes, we only consider objects located within 25~Mpc of the high resolution region centre in order to avoid contamination by low-resolution particles. At various points in this paper we consider high redshift progenitors at three snapshots: $z=10.0$, $z=13.7$, and $z=15.1$. In order to measure the median collapse times of haloes we perform the procedure below for snapshots at redshifts $z=20$ to $z=0$. 

We match $z=0$ isolated haloes with their very high redshift progenitors in the following manner. We identify the initial conditions positions for both $z=0$ and high redshift halo particles, then compute the centroid of these initial conditions positions. For each $z=0$ halo we identify the high redshift haloes whose initial conditions centroids are within 3~Mpc (comoving) of the $z=0$ halo initial conditions centroid. We use this threshold as it is approximately the comoving radius of the largest primordial proto-halo in our sample. From this selection we use the particle IDs to compute the fraction of each high redshift halo's particles that are present in the $z=0$ halo. Any high redshift halo for which the particle fraction is greater than 50~per~cent is considered a progenitor of the $z=0$ halo.

The process for identifying dwarf subhalo primary progenitors is identical to that for isolated haloes up to and including the 3~Mpc aperture for centroids. From this point, we identify a primary progenitor by computing the fraction of $z=0$ subhalo particles present in a high redshift halo. This approach is suitable for our work because we only consider dwarf ($M<10^{10}$~$\msun$) subhaloes; it would not be appropriate for more massive haloes, which would not have had the chance to collapse a sufficiently high proportion of their mass into a single progenitor at these very high redshifts.

\section{Results}
\label{sec:res}

We dedicate most of our results section to the case of isolated haloes -- Section~\ref{subsec:IsolatedHaloes} -- and end with a short discussion of subhaloes in Section~\ref{subsec:Subhaloes}. 

\subsection{Isolated haloes}
\label{subsec:IsolatedHaloes}
\begin{figure*}
    \centering
    \includegraphics[scale=0.43]{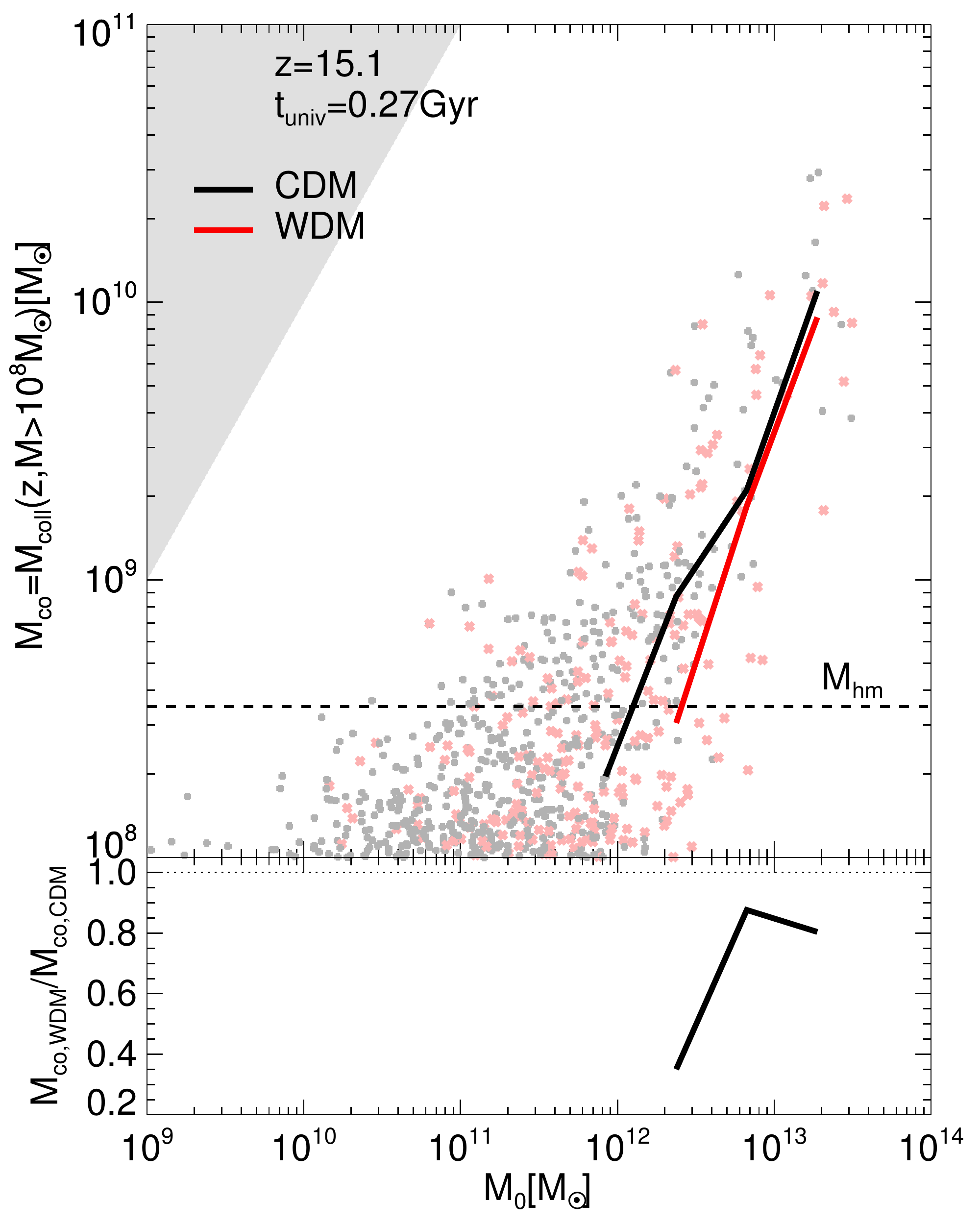}
    \includegraphics[scale=0.43]{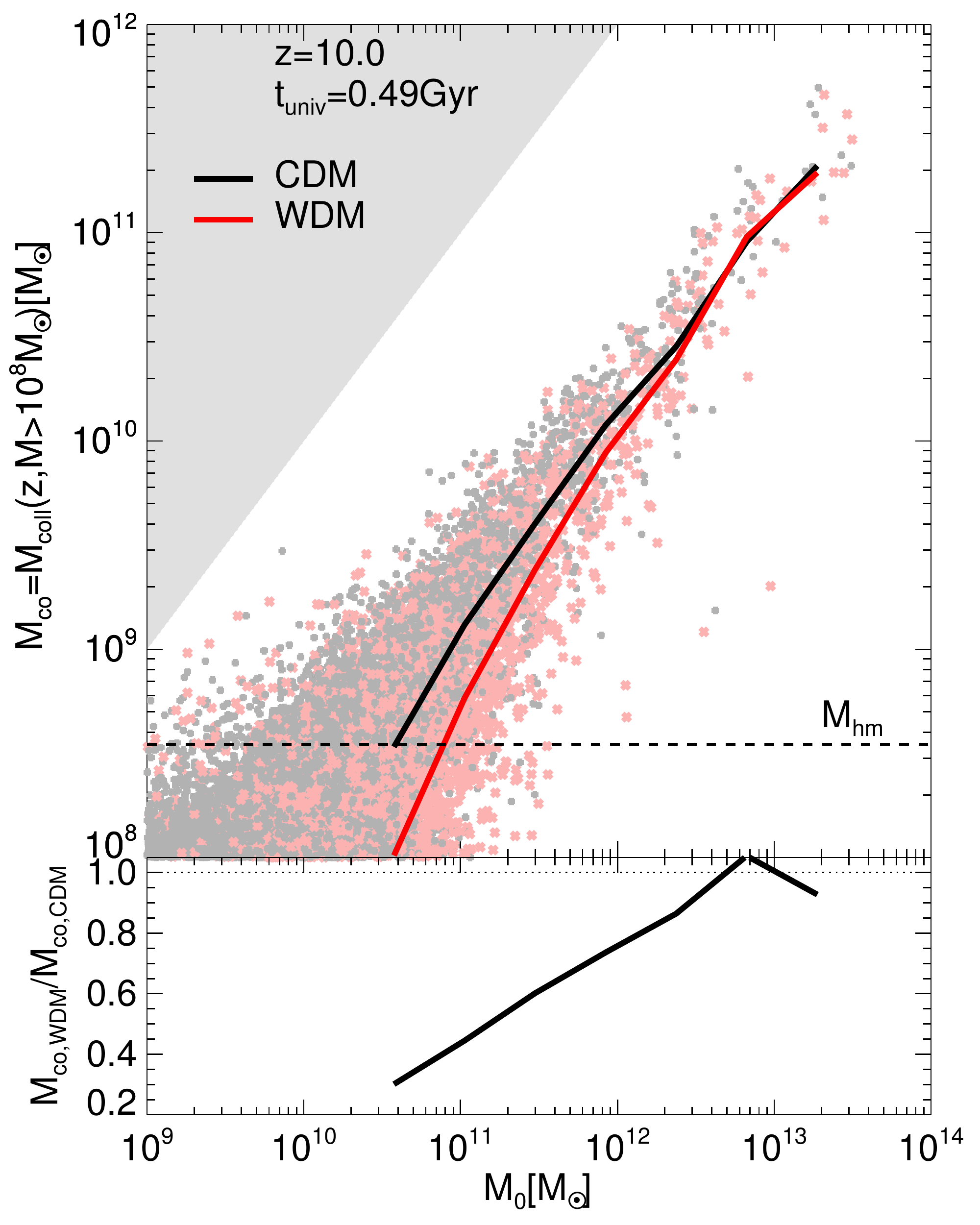}
    \caption{The collapsed mass, $M_\rmn{coll}$, as a function of halo $M_{0}$ for progenitors at $z=15.1$ (left) and $z=10.0$ (right). CDM haloes are shown as grey symbols and WDM haloes as pink symbols, with median relations plotted as black and red lines respectively. Medians are calculated using all haloes, including those for which $M_\rmn{coll}<10^{8}$~$\msun$, and therefore they do not bisect the plotted points. The half mode mass $M_\rmn{hm}$ is indicated with a horizontal dashed line.The grey triangle marks the region where $M_\rmn{coll}(z)\ge M_{0}$; note the difference in the $y$-axes between the two redshifts. We also present the ratio of the median WDM $M_\rmn{coll}$ to the median CDM $M_\rmn{coll}$ in the lower panels.}
    \label{fig:mcoll}
\end{figure*}

We begin our presentation of the results with an investigation of how much mass has collapsed for $z=0$ isolated haloes at $z\ge 10$. We identify all of the haloes -- including subhaloes -- that are high redshift progenitors of each $z=0$ isolated halo and have masses $M_\rmn{dyn}\ge10^{8}$~$\msun$, where $M_\rmn{dyn}$ is the gravitationally bound mass determined by the halo finder. We choose $M_\rmn{dyn}\ge10^{8}$~$\msun$ as this is the approximate mass threshold for atomic hydrogen cooling at $z\ge10$ \citep{Bullock_00,Benson_02,BenitezLlambay20}. We compute the sum of the progenitors' masses and refer to the result as the collapsed mass, $M_\rmn{coll}$. We perform this procedure for CDM and WDM progenitor haloes at $z=15.1$ and $z=10$, and present the results in Fig.~\ref{fig:mcoll} as a function of the $z=0$ halo virial mass, $M_{200}$, which is defined as the mass with the radius that encloses an overdensity $200\times$ the critical density for collapse. We specifically refer to the value of $M_{200}$ at the present day as $M_{0}\equiv M_{200}(z=0)$.
 
At $z=15.1$, CDM haloes of $M_{0}=10^{13}$~$\msun$ have already generated an average of $10^{10}$~$\msun$ in collapsed halo mass, or 0.1~per~cent of the final value. This drops to an average of $M_\rmn{coll}=2\times10^{8}$~$\msun$ -- $\sim500$ particles --  at $10^{12}$~$\msun$, or 0.05~per~cent; at still lower masses only a minority of haloes have any collapsed progenitors at this epoch. Remarkably, the WDM haloes have exactly the same collapsed mass fraction as CDM at $M_{0}=10^{13}$~$\msun$, before dropping to 40~per~cent of the CDM median at $2\times10^{12}$~$\msun$. Therefore, even at this very early time of 270~Myr after the Big Bang, the progenitors of galaxy group haloes are evolving in much the same way in the two models. By extension, we infer that even more massive haloes not present in the COCO volume, such as rich galaxy groups and galaxy clusters, will also exhibit identical collapsed fractions in the two models at $z\sim15$.

Moving forward 220~Myr to $z=10$, both models' galaxy group-mass haloes have attained an average $10^{11}$~$\msun$ in collapsed mass, and $\gsim 10^{10}$~$\msun$ for $M_{0}=2\times10^{12}$~$\msun$; here the ratio of WDM-to-CDM collapsed mass has increased from 40~per~cent at $z=15.1$ to 75~per~cent at $z=10$. Both models generate an average $M_\rmn{coll}>10^{8}$~$\msun$ at $3\times10^{10}$~$\msun$, which is suppressed on average by 65~per~cent in WDM relative to CDM. We detect some haloes for WDM in which the halo mass is $<M_\rmn{hm}$, even though this is nominally the mass below which no halo can form. These objects are the non-virialised central regions of haloes undergoing monolithic collapse, which are determined to be gravitationally bound by the halo finder and are therefore registered as low mass haloes. In conclusion, we demonstrate that objects with present-day mass $>10^{13}$~$\msun$ begin their formation process prior to $z=15.1$ in both CDM and WDM, and at lower $M_{0}$ there is a delay in the WDM collapse-onset time that grows larger with decreasing mass.

\begin{figure*}
    \centering
    \includegraphics[scale=0.39]{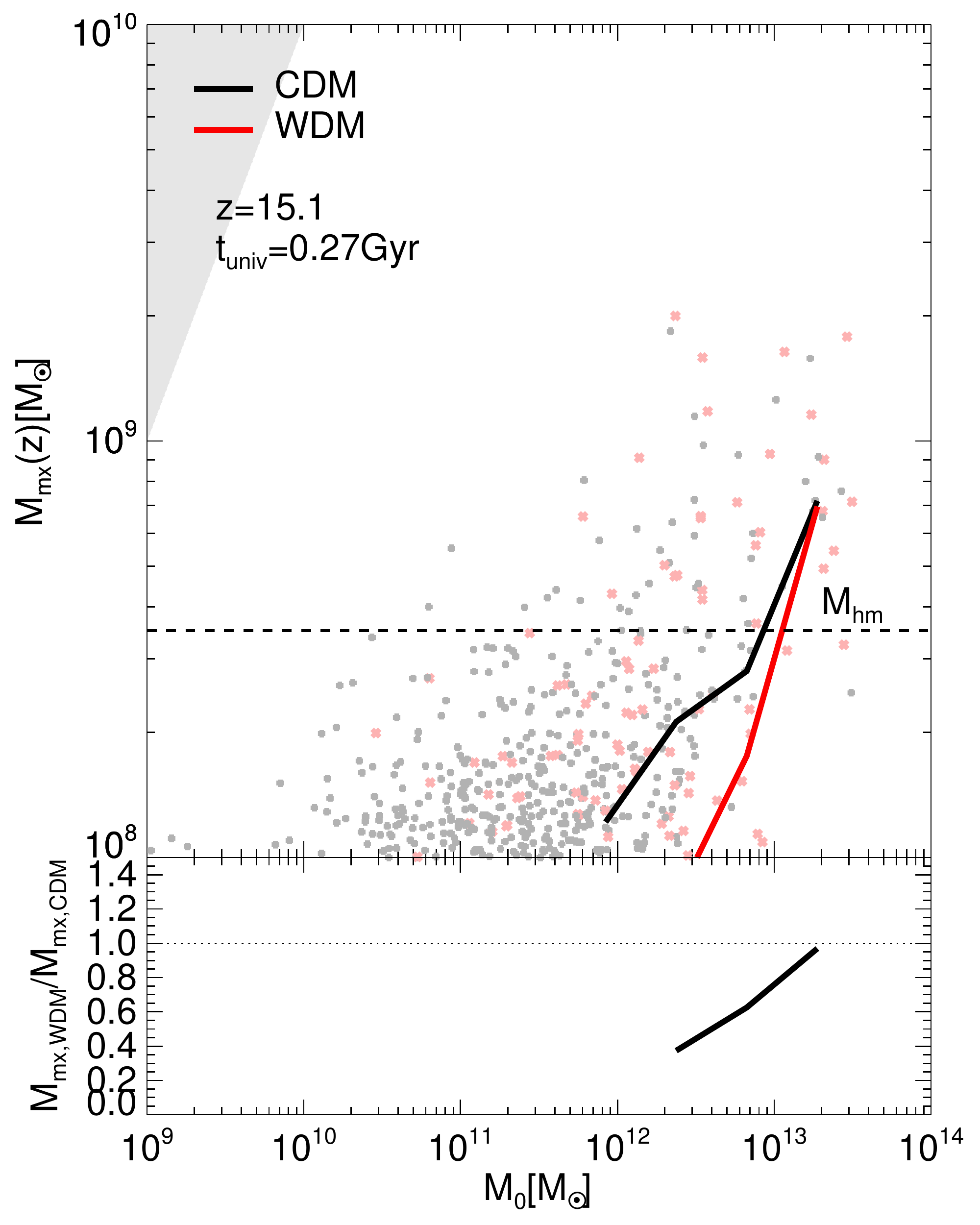}
    \includegraphics[scale=0.39]{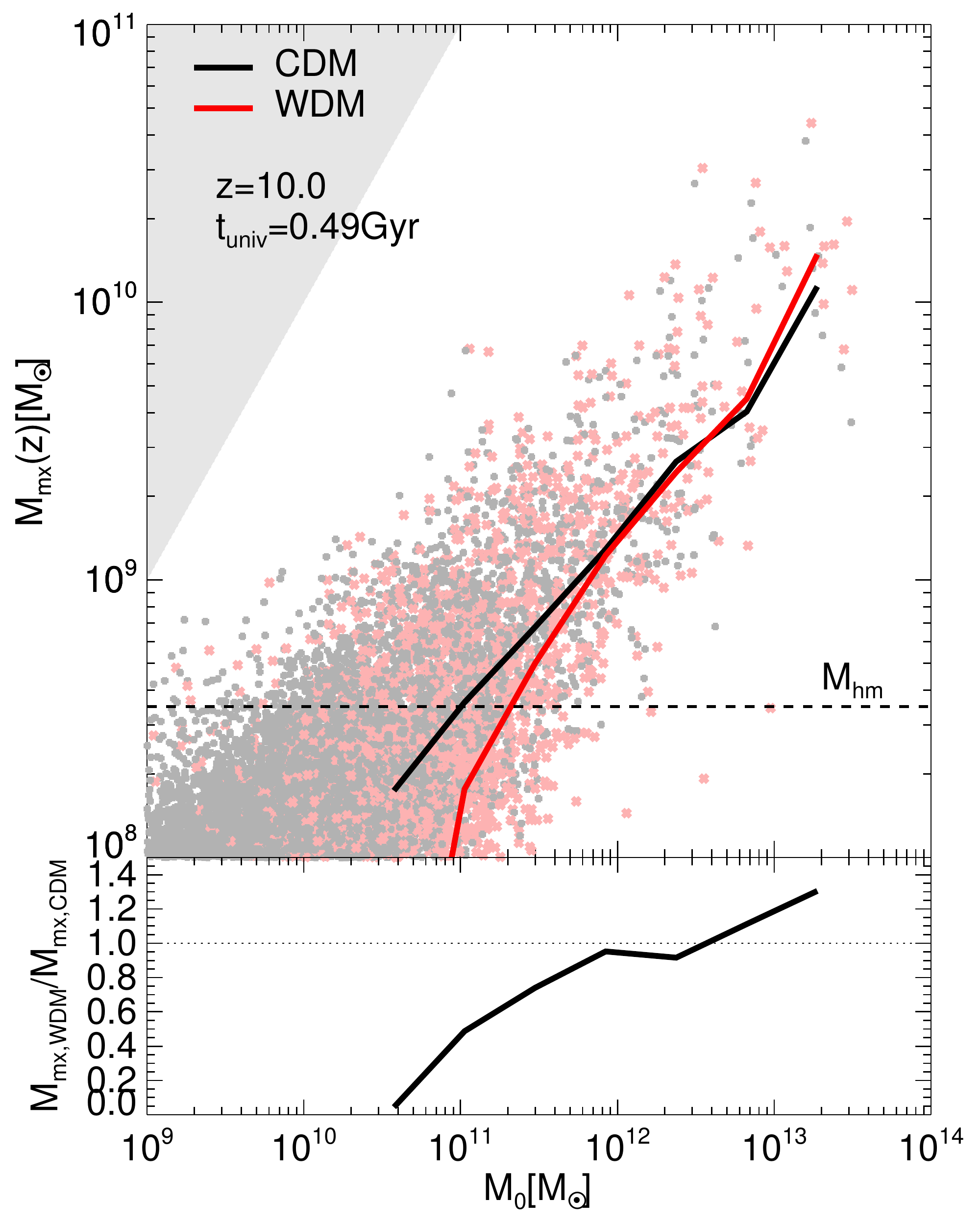}
    \caption{Most massive progenitor mass $M_\rmn{mx}$ as a function of $M_{0}$ for $z=15.1$ (left) and $z=10$ (right). Individual CDM and WDM haloes are shown as grey and pink symbols respectively, and median relations as black and red lines. Medians are calculated using all haloes, including those for which $M_\rmn{mx}<10^{8}$~$\msun$, and therefore they do not bisect the plotted points. The half mode mass $M_\rmn{hm}$ is indicated with a horizontal dashed line. The grey triangle marks the region where $M_\rmn{mx}(z)\ge M_{0}$; note the difference in the $y$-axes between the two redshifts. We present the ratio of the median values in the lower panels.}
    \label{fig:mmax}
\end{figure*}

The collapsed mass at $z>10$ is a sum over the mass of many haloes that will combine to form the final $z=0$ halo. We now turn to an alternative measure, namely the mass of the most massive progenitor at high redshift, $M_\rmn{mx}$. We define $M_\rmn{mx}$ as the highest $M_{200}$ of the available progenitors at the considered redshift. We employ $M_{200}$ here instead of $M_\rmn{dyn}$ because the latter is enhanced in WDM relative to CDM by the preference for smooth accretion over self-bound, low-mass haloes. Note that neither definition is optimal at these very high redshifts, especially for WDM. The overdensity related to virialisation may change at this epoch \citep{Eke96,Bryan98}; however, $M_{200}$ is the mass definition that is available in the halo catalogue so we shall retain this definition at all redshifts. In addition, $M_\rmn{dyn}$ estimates for WDM haloes collapsing at the cutoff scale frequently include filamentary material beyond the spherical halo \citep{Angulo13,Lovell19b}, which might not be appropriate depending on the scientific application. We plot the results for $z=15.1$ and $z=10$ in Fig.~\ref{fig:mmax}. 

The $M_\rmn{mx}$ of $z=15.1$ progenitors is $<10^{9}$~$\msun$ for both CDM and WDM, and is less then 10~per~cent of the total collapsed mass discussed in Fig.~\ref{fig:mcoll}. WDM haloes have a median $M_\rmn{mx}$ that is 60-80~per~cent of their CDM counterparts, indicating that the largest WDM haloes here are delayed in their growth relative to CDM. However, by $z=10$ WDM progenitors of present day $[10^{12},10^{13}]$~$\msun$ haloes have already caught up to the masses of CDM haloes ($[10^9,5\times10^9]$~$\msun$). Therefore, given the mass of the MW is $\sim10^{12}$ \citep{Callingham19,Deason21,Bird22}, the most massive progenitor of the MW is already in place with the same mass in WDM and CDM by $z=10$. $M_{0}=10^{11}$~$\msun$ WDM haloes are suppressed to 60~per~cent of the $M_\rmn{mx}$ of CDM counterparts, which is a smaller suppression than the total collapsed mass at the same mass. This result reflects the delay in the formation of less massive progenitors and potentially the suppression of the WDM halo mass function. In summary, we have developed a picture in which both CDM and WDM galaxy group mass haloes start forming  at $z\sim15$; the collapse of $10^{12}$~$\msun$ haloes lags behind in WDM, but by $z=10$ the most massive progenitor rapidly catches up with a small number of lower mass. Lower mass haloes are delayed still further in their collapse.

\begin{figure*}
    \centering
    \includegraphics[scale=0.34]{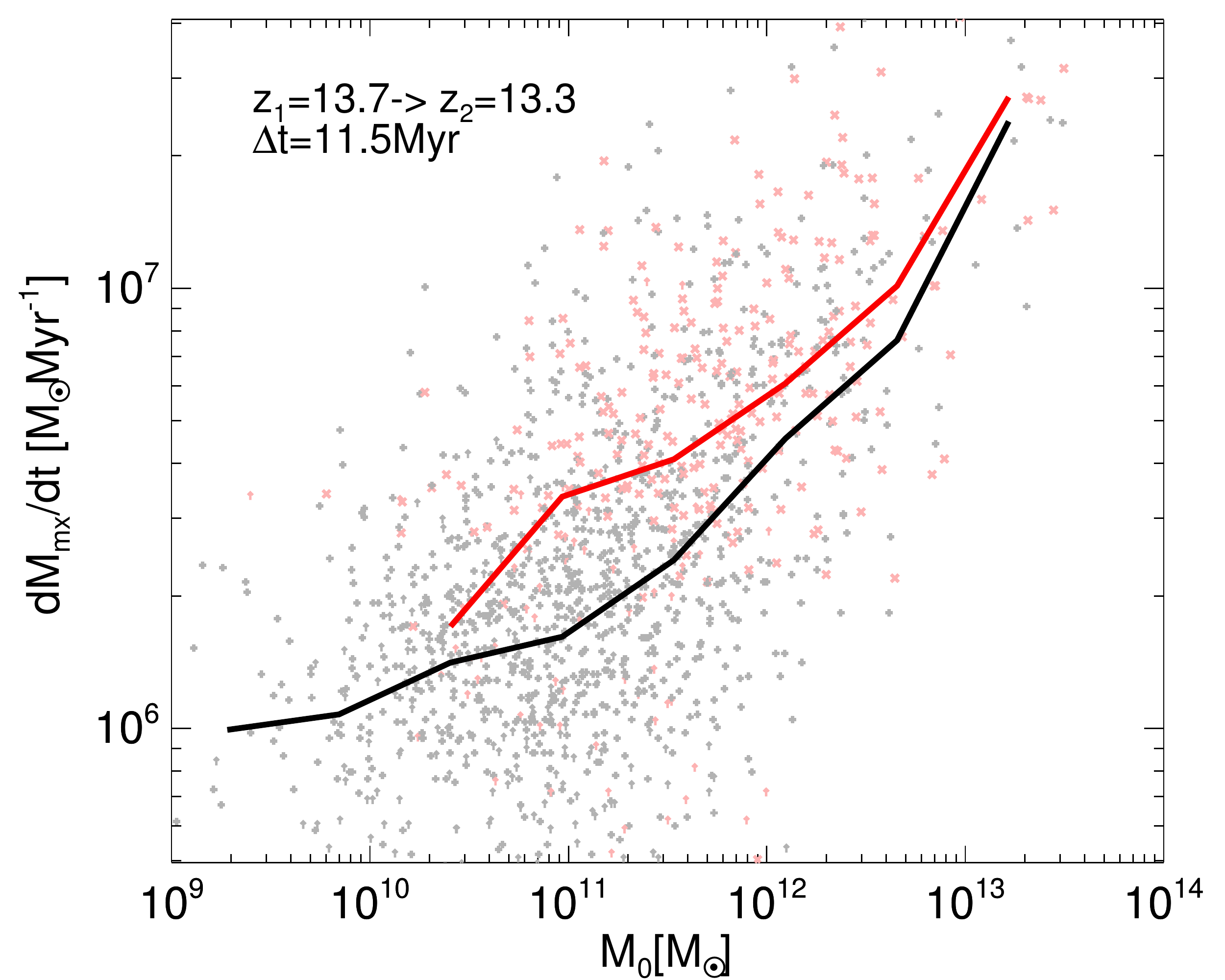}
      \includegraphics[scale=0.34]{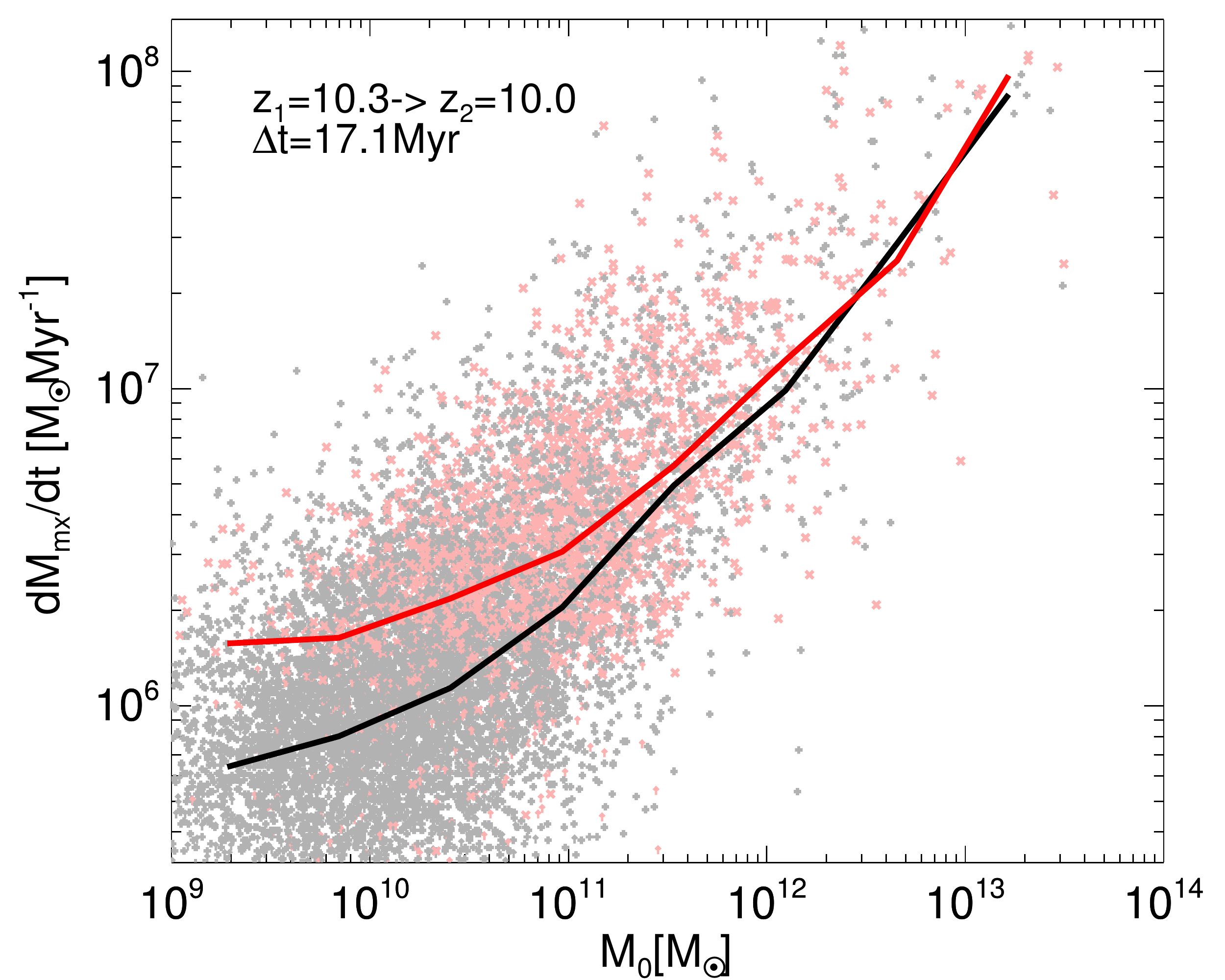}
    \caption{The growth rate in $M_\rmn{mx}$ at $z=13.3$ (left-hand panel) and $z=10.0$ (right-hand panel) as a function of $M_{0}$. CDM data are shown as grey symbols, and WDM as pink symbols, median relations are indicated with black and red lines respectively. Haloes for which $M_\rmn{mx}<10^{8}$~$\msun$ at the previous snapshot are shown as arrows to mark lower limits.}
    \label{fig:mcollev}
\end{figure*}

The rapid growth of structure in WDM is of interest due to the potential for bright starbursts \citep{Governato15,Bose16c,Lovell19b}, which could plausibly be detectable and may have a significant impact on galaxy formation. We investigate this phenomenon by computing the change in collapsed mass for each object at $z=13.3$ and $z=10$, which we measure as the ratio increase in $M_\rmn{mx}$ from the previous snapshot. We use $z=13.3$ rather than $z=15.1$ due to the combination of the lack of objects at $z=15.1$ and the difficulty in measuring growth rates as opposed to total mass. For haloes that have a measurable collapsed mass at the previous snapshot, we compute the difference in $M_\rmn{mx}$ between these two redshifts divided by the time difference as a function of present day halo mass; in cases where a collapsed mass cannot be measured at the previous snapshot we instead compute a lower limit by subtracting the minimum permitted mass, $10^{8}$~$\msun$. We present the results in Fig.~\ref{fig:mcollev}; note that we only include haloes for which $M_\rmn{mx}$ is large enough to be measured at $z=13.3$.  

The CDM haloes' average growth rate per unit time at $z=13.3$ increases steadily from $1.3\times10^{6}$~$\msun\rmn{Myr}^{-1}$ at $M_{0}=10^{11}$~$\msun$ to $2\times10^7$~$\msun\rmn{Myr}^{-1}$ at $2\times10^{13}$~$\msun$. A similar increase with halo mass occurs for $z=10.0$, from $2\times10^{6}$~$\msun\rmn{Myr}^{-1}$ at $10^{10}$~$\msun$ to $8\times10^{7}$~$\msun\rmn{Myr}^{-1}$ at $2\times10^{13}$~$\msun$. As the mass decreases further below $10^{10}$~$\msun$ the change in the growth rate becomes smaller, which is a result of our choice to only include haloes for which $M_\rmn{mx}>10^{8}$~$\msun$ at each redshift: including less massive haloes would in principle bring down the growth rate, although this cannot be measured reliably at our available resolution. The WDM model behaves very similarly to CDM at $M_{0}$ greater than some threshold -- $6\times10^{12}$~$\msun$ at $z=13.3$, and $4\times10^{12}$~$\msun$ at $z=10$ notwithstanding some statistical noise -- followed by an excess growth rate of up to $3\times$ the corresponding CDM value. We can expect that this growth rate will generate intense starbursts, which may make for very bright objects in WDM that are not present in CDM; modelling this effect requires careful attention to gas accretion and stochastic simulation effects, and therefore we defer this analysis to future work. The delay in collapse is indicated by the lack of measured growth rates in WDM for $M_{0}<10^{11}$~$\msun$ at $z=13.3$. We therefore retrieve the familiar picture in which WDM halo formation is delayed relative to CDM. When WDM halo collapse finally does take hold the growth rate is around three times the CDM rate; it soon catches up and subsequently follows the CDM mass accretion history. 

\begin{figure}
    \centering
    \includegraphics[scale=0.335]{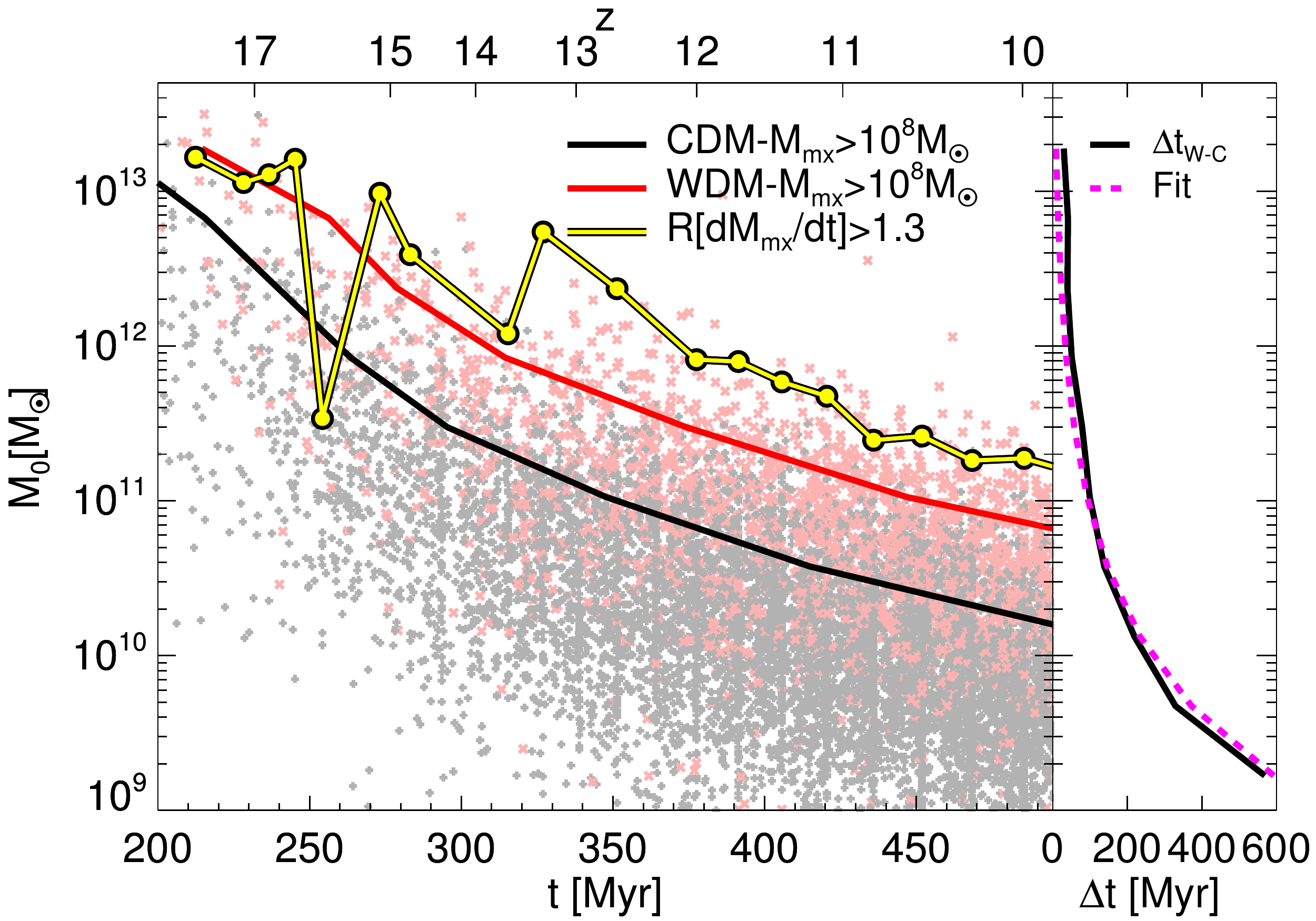}     
    \caption{The relationship between the time at which a halo first reaches $M_\rmn{mx}=10^{8}$~$\msun$ and its present day virial mass. Individual haloes are shown with grey (pink) symbols for CDM (WDM), and the median times of first $M_\rmn{mx}=10^{8}$~$\msun$ as a function of $M_{0}$ are shown as black and red lines respectively. The yellow line indicates the $M_{0}$ for which the ratio of the median WDM and CDM growth rates is $<1.3$. In the right hand panel we show the difference in the median time to $M_\rmn{mx}=10^{8}$~$\msun$ between WDM and CDM as a black line. We also indicate a fit to this delay time-$M_{0}$ relation with the dashed magenta line.}
    \label{fig:delmmax}
\end{figure}

We have shown that WDM haloes have delayed onset of collapse relative to CDM, followed by a time of accelerated collapse after which halo growth proceeds at the same pace in both models. Furthermore, we show that the delay in collapse onset is influenced by the size of the overdensity mass relative to the half-mode mass. We now summarise this behaviour with a further measure of collapse time: the time at which each present-day halo first exhibits a progenitor with $M_\rmn{mx}=10^{8}$~$\msun$. We denote this time $t_{8}$. This mass is of interest both as the approximate atomic cooling limit at high redshift and in its proximity to $M_\rmn{hm}$ for this WDM model. In Fig.~\ref{fig:delmmax} we present the $M_{0}$--$t_{8}$ values for both cosmologies, along with the median $t_{8}$ as a function of $M_{0}$ in the time frame $t=[200,500]$~Myr, or approximately from $z=18$ to $z=10$. Many haloes with $M_{0}<10^{12}$~$\msun$ attain the $10^{8}$~$\msun$ threshold at $z<10$, therefore we use data through to $z=0$ to calculate the medians. These relations mark the approximate onset of galaxy formation and, in the case of WDM, the start of monolithic collapse. We also indicate the regime in which halo growth is approximately the same in CDM and WDM by computing the ratio of the halo growth rate between the two models and identifying the lowest value of $M_{0}$ at which this ratio takes the value $1.3$. We adopt this value of $1.3$ to be the point where growth in WDM has approximately caught up to CDM as at smaller values the ratio becomes noisy.     

CDM haloes have already attained $M_\rmn{mx}>10^{8}$~$\msun$ at $z>17$ for galaxy group mass haloes, and by the end of the notional {\it JWST frontier epoch} at $z=10$ large dwarfs ($M_{0}\sim2\times10^{10}$~$\msun$) have done likewise, albeit with considerable scatter. WDM group haloes also start their collapse before $z>17$, delayed on average by $\sim20$~Myr relative to their CDM counterparts. However, the delay in collapse time grows with $M_{0}$ as anticipated from the argument concerning the lack of high density peaks in WDM, such that by $z=10$ it is only haloes of $M_{0}>6\times10^{10}$~$\msun$ that have collapsed, some 100~Myr later than CDM haloes of the same mass. We indicate the difference in the median collapse times in the right-hand panel of Fig.~\ref{fig:delmmax}, which shows the increase in delay time from $\sim20$~Myr at $M_{0}=2\times10^{13}$~$\msun$ to 300~Myr at $5\times10^{9}$~Myr. We approximate this relation with an equation of the form:

\begin{equation}
    \Delta t_{8} = 1.2~\rmn{Gyr}\cdot\left(\frac{M_\rmn{hm}}{M_{0
}}\right)^{0.45}.
\end{equation}

Note that we have only fit these equation parameters for $M_{0}>2\times10^{9}$~$\msun$. In principle, one can attempt to link the peak overdensity $\delta_{0}$, to $M_{0}$ in the two models and then use the measured $\Delta t$ to calibrate equation~\ref{eqn:td0}; we defer this type of analysis to future work. This relation predicts that the delay in collapse time of an independent halo of mass $M_{0}=M_\rmn{hm}$ is 1.2~Gyr. We will demonstrate below that the delay is shorter for subhaloes, and therefore 1.2~Gyr can be thought of as a maximum delay time given that very few haloes form with $M_{0}<M_\rmn{hm}$ in WDM.

Our choice of time where WDM growth catches up to the CDM growth rate, as indicated by the yellow line, roughly traces the upper limit of the WDM $t_{8}$-$M_{0}$ values, although the calculation is beset by noise issues. It suggests that CDM and WDM behave the same for $M_{0}>10^{12}$~$\msun$ at $z\sim13$ and $M_{0}>2\times10^{11}$~$\msun$  at $z\sim10$. In conclusion, we have demonstrated that the initial collapse of a halo is delayed in WDM with respect to CDM by a factor $\propto M_{0}^{-0.45}$, and that by $z=10$ all haloes with $M_{0}>2\times10^{11}$~$\msun$ are growing at the same rate in the two models. 

\subsection{Subhaloes}
\label{subsec:Subhaloes}
\begin{figure*}
    \centering
    \includegraphics[scale=0.35]{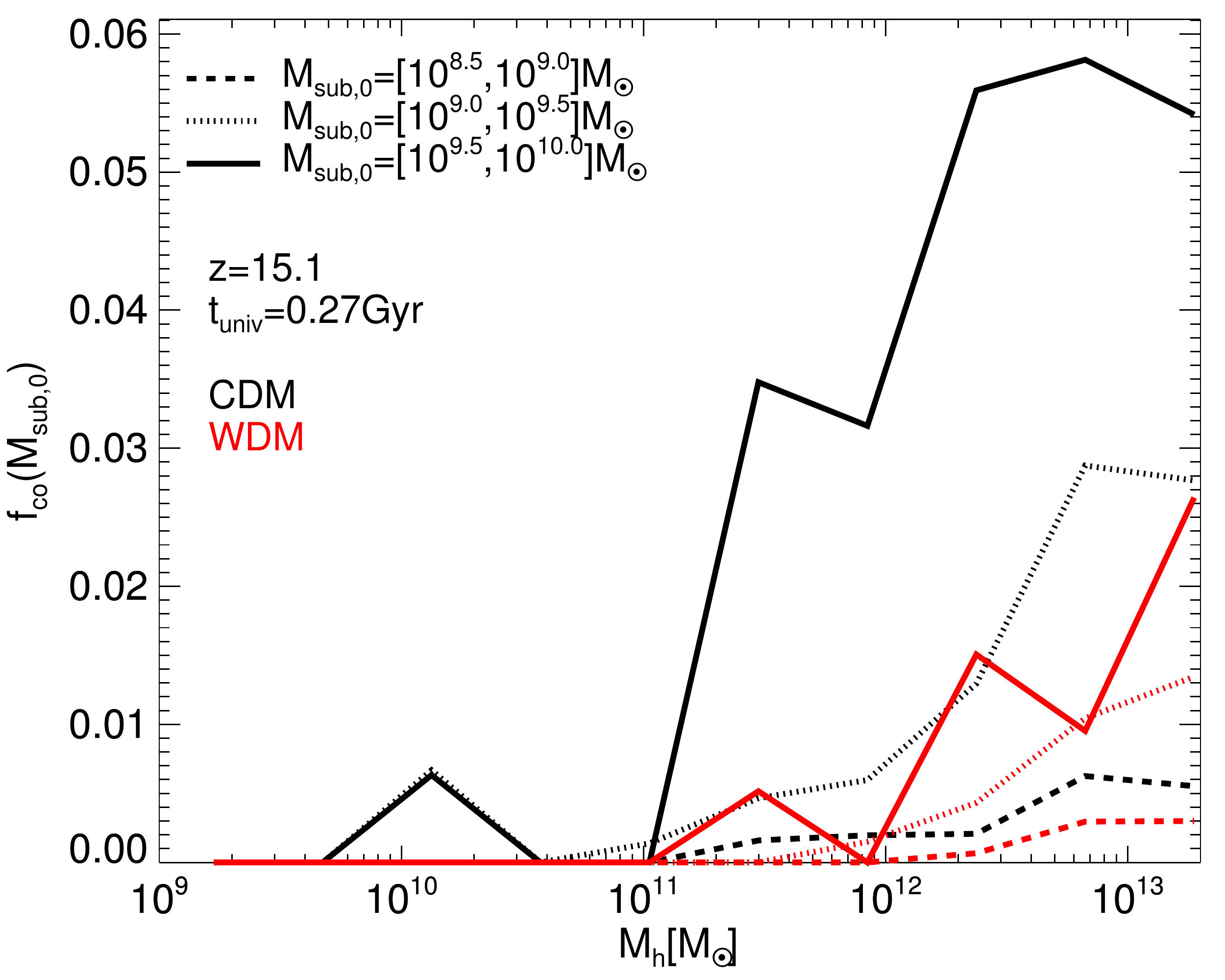}
    \includegraphics[scale=0.35]{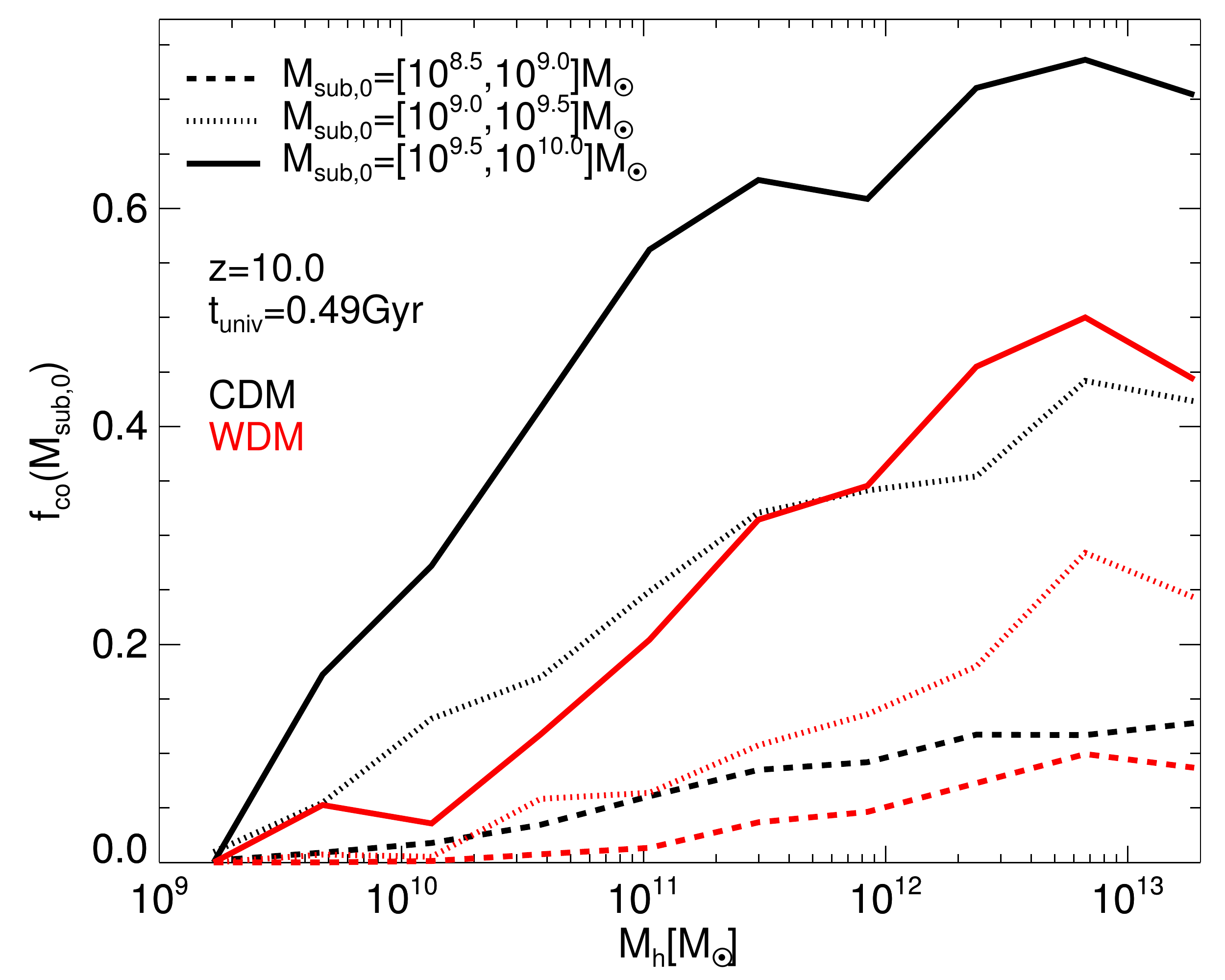}
    \caption{Subhalo collapsed fractions as a function of $z=0$ host halo mass, $M_\rmn{h}$, measured at redshifts $z=15.1$ (left) and $z=10.0$ (right). We present collapsed fractions for subhaloes in the mass brackets $M_\rmn{sub,0}=[10^{8.5},10^{9.0}]$~$\msun$, $M_\rmn{sub,0}=[10^{9.0},10^{9.5}]$~$\msun$ and $M_\rmn{sub,0}=[10^{9.5},10^{10.0}]$~$\msun$ as dashed, dotted, and solid lines respectively. CDM data are shown in black and WDM data in red. Note the difference in the $y$-axis values between the two panels.} 
    \label{fig:msub}
\end{figure*}

We end our presentation of the results with a consideration of the assembly history of subhaloes, in the context of the host haloes within which they reside at $z=0$. The density of the Universe at the subhalo collapse time sets the subhalo's central density, with earlier collapse times leading to higher central densities. Given the increase in $\delta_{0}$ associated with being in close proximity to the density perturbation of a future massive halo, we therefore expect that subhalo collapse times, and by extension densities, will correlate with host mass. The subhaloes of massive hosts will form first, and then lower mass hosts' subhaloes will collapse later. Using the method described in Section~\ref{sec:sims} we identify the fraction of subhaloes that have condensed at $z=15.1$ and $z=10$ as a function of the host halo $M_{0}$ -- which we label $M_\rmn{h}$ -- for subhaloes in three $z=0$ mass bins: $M_\rmn{sub,0}=[10^{8.5},10^{9.0}]$~$\msun$, $M_\rmn{sub,0}=[10^{9.0},10^{9.5}]$~$\msun$ and $M_\rmn{sub,0}=[10^{9.5},10^{10.0}]$~$\msun$, where $M_\rmn{sub,0}\equiv M_\rmn{dyn}(z=0)$; we label this collapsed fraction $f_\rmn{co}$. We present the results for CDM and WDM at these two redshifts and in these three $M_\rmn{sub,0}$ bins in Fig.~\ref{fig:msub}.    

In all cases $f_\rmn{co}$ increases monotonically with host mass up to $M_\rmn{h}=7\times10^{12}$~$\msun$, reflecting how the increase in the overdensity amplitude around more massive perturbations leads to the earlier formation of its future subhaloes. $f_\rmn{co}$ also increases with $M_\rmn{sub,0}$, reflecting the larger contribution to $\delta_{0}$ from these haloes in and of themselves. 

The highest median value of $f_\rmn{co}$ for any subhalo mass bin at $z=15.1$ is 6~per~cent, which occurs for the highest mass CDM subhalo bin in $M_\rmn{h}=6\times10^{12}$~$\msun$ haloes. WDM haloes in the same bin are suppressed by a factor of six and are lower in their collapsed fractions than even the CDM $M_\rmn{sub,0}=[10^{9.0},10^{9.5}]$~$\msun$ subhaloes. By $z=10$ the highest mass bin haloes have increased by 0.69 in CDM compared to by $\sim0.48$ in WDM, thus unlike the main halo progenitors a lag remains in WDM for subhaloes. This phenomenon is also apparent in the gradient of the $z=10$ collapsed fractions between $M_\rmn{h}=10^{11}$~$\msun$ and $M_\rmn{h}=6\times10^{12}$~$\msun$, which is steeper for WDM in all three mass bins.  

We have also checked for the impact of peak mass on these relations. We computed the ratio of the peak value of $M_{200}$ for each subhalo prior to accretion to the $M_\rmn{sub,0}$. We compute the median ratio for each of the Fig.~\ref{fig:msub} host mass and subhalo mass bins at $z=10.0$; we avoid $z=15.1$ due to poor statistics for this metric.  For low mass hosts -- $4\times10^{10}$~$\msun$ -- all three mass bins for both models show that the median peak mass was 70~per~cent higher than the present day $M_\rmn{sub,0}$. The ratio then drops with increasing host mass, indicating that these subhaloes have undergone less stripping, and the drop is steeper for lower mass haloes. This result is expected due to the stripping rate being related to the ratio of pre-infall mass to host mass. In the context of Fig.~\ref{fig:msub}, we conclude that the $f_\rmn{co}$--$M_\rmn{h}$ is slightly shallower than it would be if we were binning subhaloes by $M_\rmn{peak}$ rather than by $M_\rmn{sub,0}$. We find that this method appears to show a few more per~cent of stripping in CDM than in WDM. We attribute this result to be an artifact of the fact that $M_\rmn{sub,0}$ does not include substructure, which forms a greater fraction of the mass in CDM than in WDM and so the ratio of the peak mass -- which includes substructure -- to $M_\rmn{sub,0}$ will be bigger in CDM.  

\section{Conclusions}
\label{sec:conc}

The collapse of structure contains vital information about the nature of dark matter, and in particular the erasure of small scale perturbations by free-streaming in warm dark matter (WDM) has been shown previously to delay the onset of collapse. In this paper we show how the assembly history changes at very early times, $z>10$, to obtain an understanding of how previous studies have shown a remarkably small difference between the expected number of galaxy detections with {\it JWST} despite this delay \citep{Bose16c,Menci16,Lovell19b,Kurmus22,Maio23}. We consider the result of \citet{Eke96} that the collapse time is strongly dependent on the total overdensity associated with a dark matter halo prior to collapse, and therefore the delay correlates with $z=0$ descendant mass: massive descendants (virial mass $M_{0}>10^{12}$~$\msun$) have large overdensities that are affected minimally by the WDM cutoff, and as the descendant mass decreases the impact of overdensity erasure by free-streaming increases, leading to progressively longer delays in collapse.

We approach this problem by selecting CDM and WDM (3.3~keV thermal relic) haloes from the COCO $N$-body simulations, both isolated haloes with present day virial masses in the range $[10^{8.5},10^{13.5}]$~$\msun$ and subhaloes of dynamical mass $[10^{8.5},10^{10.0}]$~$\msun$. We identify progenitors to these haloes at $z=10$ and $z=15.1$, which we term the {\it JWST frontier epoch}, compute the total mass collapsed into haloes of dynamical mass $>10^{8.0}$~$\msun$ and identify the most massive progenitor, and use the results to examine how halo assembly proceeds differently for different mass $z=0$ progenitors in CDM and WDM.

We show that the collapsed mass, $M_\rmn{co}$, in group mass haloes -- $M_{0}=10^{13}$~$\msun$ -- is already nearly identical between CDM and WDM at $z=15.1$ with an average of $M_\rmn{co}=10^{10}$~$\msun$ (Fig.~\ref{fig:mcoll}). The difference is larger at $2\times10^{12}$~$\msun$, but the gap at this mass closes from a WDM suppression of 40~per~cent relative to CDM at $z=15.1$ to 20~per~cent by $z=10$, thus corroborating the statement that larger perturbations exhibit only small differences in collapse time between CDM and WDM, and the WDM delay grows for progressively less massive descendants. We find even smaller differences for the most massive progenitor at these two redshifts (Fig.~\ref{fig:mmax}). 

The growth rate of haloes will therefore differ in order for WDM to catch up to CDM. We demonstrate that the growth rate of WDM haloes is higher than that of CDM for all considered masses and redshifts at which a growth rate can be measured (Fig.~\ref{fig:mcollev}), which is $M_{0}>10^{9}$~$\msun$ at $z=10$ and $M_{0}>10^{11}$~$\msun$ at $z=13.3$. Finally, we show that present day subhaloes collapse times are strongly dependent on both their present day mass and their host mass, with earlier collapse times both both increasing subhalo mass and increasing host mass (Fig.~\ref{fig:msub}). The WDM subhaloes collapse later than their CDM counterparts, in common with their host halo counterparts. The slope of the collapse fraction-descendant host mass relation is steeper in WDM than in CDM, indicating that the collapse delay is larger for subhaloes in low mass hosts than in large mass hosts, reflecting our initial hypothesis concerning the relationship between collapse time and the total overdensity $\delta_{0}$.

The primary consequence of this work is that we illustrate the challenge for high redshift studies in testing predictions for different dark matter models. We have shown that the cutoff-induced delay in collapse around $z=10$ is only discernible for galaxies whose descendants are of a mass $M_{0}<8\times10^{11}$~$\msun$, or just a few times more massive than the Large Magellanic Cloud. Very high quality lensing data may be needed to detect this class of object \citet{Shen23}. The best option may be to attempt to find the first satellites of these high redshift objects in the hope of detecting faint starbursts during monolithic collapse.

We can also draw conclusions for further studies of faint Local Group systems. The easiest objects to simulate and study in large quantities are isolated dwarfs that are distant from massive galaxies, whereas the easiest observed galaxies to study observationally are Milky Way (MW) satellites, which will be influenced by the MW's own density perturbation. We have shown that the influence of free-streaming is stronger on isolated dwarfs than on satellites of massive galaxies, therefore one must interpret studies of isolated galaxies as providing an extreme limit on the possible deviation of MW satellite properties between CDM and WDM.

\section*{Acknowledgements}

MRL acknowledges support by a Grant of Excellence from the Icelandic Research Fund (grant number 206930). We would like to thank Wojtek Hellwing, Jes\'us Zavala and Tamar Meshveliani for helpful comments on the text.

\section*{Data Availability}

The COCO-CDM and COCO-WDM simulations were originally published in  \citet{Hellwing16} and \cite{Bose16a}. Requests for access should contact Wojtek Hellwing.



\bibliographystyle{mnras}








\bsp	
\label{lastpage}
\end{document}